# Decentralized Health Intelligence Network (DHIN)

Abraham Nash
University of Oxford
abraham.nash@cs.ox.ac.uk

**Abstract:** *Decentralized Health Intelligence Network (DHIN)* extends the *Decentralized Intelligence Network (DIN)* framework to address challenges in healthcare data sovereignty and AI utilization. Building upon *DIN's* core principles, *DHIN* introduces healthcare-specific components to tackle data fragmentation across providers and institutions, establishing a sovereign architecture for healthcare provision. It facilitates effective AI utilization by overcoming barriers to accessing diverse health data sources. This comprehensive framework leverages: 1) self-sovereign identity architecture coupled with a personal health record (PHR), extending DIN's personal data stores concept to ensure health data sovereignty; 2) a scalable federated learning (FL) protocol implemented on a public blockchain for decentralized AI training in healthcare, tailored for medical data; and 3) a scalable, trustless rewards mechanism adapted from *DIN* to incentivize participation in healthcare AI development. *DHIN* operates on a public blockchain with an immutable record, ensuring that no entity can control access to health data or determine financial benefits. It supports effective AI training while allowing patients to maintain control over their health data, benefit financially, and contribute to a decentralized ecosystem. Unique to *DHIN*, patients receive rewards in digital wallets as an incentive to opt into the FL protocol, with a long-term roadmap to fund decentralized insurance solutions. This approach introduces a novel, self-financed healthcare model that adapts to individual needs, complements existing systems, and redefines universal coverage, showcasing how *DIN* principles can transform healthcare data management and AI utilization while empowering patients.

# 1. Introduction

Healthcare is undergoing a profound digital transformation, propelled by the widespread adoption of electronic health records (EHRs) and the growing potential of artificial intelligence (AI) in medical diagnostics and treatment [1], [2]. However, this shift faces significant challenges in data management, privacy, and interoperability. Fragmentation and siloing of patient health data across various providers and institutions not only undermine data sovereignty but also hinder the effective utilization of AI in healthcare.

To address these challenges, this paper proposes *Decentralized Health Intelligence Network* (*DHIN*), an innovative framework that builds upon the principles of *Decentralized Intelligence Network* (*DIN*) [3]. *DHIN* aims to redefine healthcare data management by leveraging blockchain technology, federated learning, and cryptographic techniques.

**This comprehensive approach encompasses:**

1. A sovereign architecture for healthcare provision, utilizing self-sovereign identity and personal health records.
2. A scalable federated learning protocol implemented on a public blockchain, enabling decentralized AI training while preserving data privacy.
3. A trustless rewards mechanism to incentivize participation and ensure fair compensation for data contributions.

By addressing the critical issues of data sovereignty, AI utilization, and patient empowerment, *DHIN* offers a promising solution to the current limitations in healthcare data management. This framework not only supports effective AI training but also introduces a novel, sustainably financed healthcare model that adapts to individual health needs.

The structure of this paper is as follows: **Section 2** presents a detailed problem statement, outlining the current challenges in healthcare data management. **Section 3** establishes the specific requirements of the *DHIN* framework. **Section 4** provides an overview of the systems architecture, detailing the root, real-world, and intelligence layers. **Section 5** delves into the methodology, explaining self-sovereign server technology, personal health records, and the decentralized federated learning protocol. **Section 6** discusses the threat model, addressing potential security concerns. **Section 7** explores the concept of a decentralized insurance solution. Finally, **Section 8** onwards concludes the paper, suggesting future research directions and discussing the broader implications of this work.

## 2. Problem Statement

The current healthcare landscape is marked by significant fragmentation and siloing of patient health data across various providers and institutions. This fragmentation undermines data sovereignty and hinders the effective utilization of artificial intelligence (AI) in healthcare. The ramifications are severe:

1. Patient identification errors contribute to approximately 195,000 deaths per year in the US alone [4].
2. Inefficient and error-prone data collection processes create fatigue for patients and healthcare providers [4].
3. Electronic Health Records (EHR) are often stored in siloed databases, becoming targets for security breaches [5], [6].
4. Lack of EHR interoperability impedes AI practitioners from developing and implementing diagnostic tools and decision-support systems [7].
5. Patients are excluded from the ecosystem of valued health data exchange, lacking ownership and control over their personal health information [8], [9].

These issues collectively obstruct the delivery of efficient, safe, and high-quality healthcare, hinder medical research advancement, and impede the development of innovative diagnostic tools and treatment pathways.

## 3. Requirements

***DHIN* adapts the core tenets of *DIN* to the healthcare domain, specifically addressing the requirements that:**

- No authority can resume access and management controls over patients' data.
- Only patients can decide who accesses their data for federated learning.
- No third-party broker determines reward allocations for patients' contributions.

*DIN* is a cross-sector framework, originally designed to preserve participant sovereignty over data while fostering collaborative AI efforts across various domains [3]. In the context of *DHIN*, this generic framework is repurposed to address the unique challenges and opportunities within healthcare.

This healthcare-specific adaptation ensures that no single authority controls the FL process, preserving patient sovereignty over their health data while fostering collaborative AI efforts. Crucially, the *DHIN* framework prevents data other than model updates from needing to leave the Personal Data Stores (PDS), maintaining user privacy and control in a healthcare context.

*Decentralized Health Intelligence Network* (*DHIN*), as a healthcare application use case of *DIN*, acknowledges the ongoing existence of institutional silos in traditional healthcare. Designed to facilitate a transition towards decentralized, sovereign data stores controlled by individuals, *DHIN* offers flexible integration with Electronic Health Records (EHR) and institutional data systems, supporting a transition to decentralized, sovereign data stores. By allowing broad participation in the federated learning (FL) protocol, the *DHIN* protocol complements new avenues for access to scalable data for AI engineering in a decentralized fashion, specifically within the healthcare sector.

While enhancing data access, privacy, security, and monetization, *DHIN* acknowledges that institutional silos and centralized learning are likely to continue, enabling a truly global reach and inclusivity, fostering a decentralized and sovereign AI development landscape within the healthcare sector. The self-sovereign identity layer may address ongoing complexities such as healthcare custodianship and emergency care, allowing *DHIN* to either operate independently or integrate with existing systems; however, this discussion extends beyond the scope of this paper. This paper illustrates *DIN*'s adaptability to healthcare data sovereignty and AI challenges while promoting a fair and secure environment for AI development that respects individual rights and encourages standardization of data formats.

## 4. Systems Architecture: An Overview

This framework supports robust and ethical AI training in healthcare by allowing patients to maintain sovereignty over their health data while simultaneously benefiting financially and contributing to a decentralized, scalable ecosystem. This ecosystem harnesses collective AI capabilities to advance healthcare algorithms, demonstrating new possibilities in medical research and patient care.

**To meet this framework's requirements and overcome the challenges specific to healthcare, *DHIN* leverages three key components:**

1. **Root Layer:** This foundational layer provides an overview of self-sovereign server technology (SSST), which underpins the system's ability to maintain individual data sovereignty in healthcare ecosystems.
2. **Real-World Layer:** This layer focuses on practical healthcare workflows—such as prescriptions, referrals, and investigations—conducted in a sovereign manner within the *Decentralized Healthcare Intelligence Network* (*DHIN*). It integrates principles from existing frameworks to ensure secure, patient-controlled exchanges. Additionally, the layer emphasizes the flow of information through the sovereign integration of Medical AI Devices, which interact with patient-owned health records. This enhances healthcare delivery by enabling efficient, transparent, and patient-centric care within the *DHIN* framework.
3. **Intelligence Layer:** This layer outlines the integration of the *DIN* protocol for a scalable and secure trustless federated learning process. It also explains the specific utility of the reward mechanism in funding decentralized healthcare insurance solutions.

By leveraging these decentralized mechanisms across these three layers, the *DHIN* framework not only addresses the current challenges in health data management but also catalyzes the adoption of scalable, sovereign data solutions. These solutions uphold individual rights, foster technological advancements, and pave the way for new use cases in healthcare. Ultimately, *DHIN* supports the ongoing transition towards decentralized data ownership, promising more equitable, efficient, and innovative healthcare ecosystems for the future.

**Key stakeholders previously specified in the *Decentralized Intelligence Network* (DIN) ecosystem include [3]:**

- **Patients:** Individuals who own and control their health data stores. They contribute data to the federated learning process while maintaining privacy and benefiting from collaborative AI training.
- **Model Owners:** Entities such as pharmaceutical companies, research institutions, or AI developers who utilize the FL protocols to enhance their models with decentralized health data, without compromising individual data sovereignty.
- **Evaluators:** Network-staked entities responsible for decentralized auditing. They ensure transparency and fairness in evaluating patients contributions and overseeing the distribution of rewards.

**Building upon these three stakeholders introduced in the *DIN*, *DHIN* introduces two additional key ecosystem stakeholders:**

- **Healthcare Professional:** Healthcare professionals (i.e., a physician) who use self-sovereign server technologies (SSST) to securely interact with the network, verify patient relationships, and access health records while maintaining high standards of privacy and security.
- **Medical AI Devices:** AI-powered tools and devices with their own public/private key identifiers, capable of interacting with patient data to support diagnosis, treatment, and healthcare delivery within the decentralized ecosystem.

This expanded set of stakeholders enhances *DHIN's* capability to address healthcare-specific challenges and opportunities, fostering a more comprehensive and efficient decentralized health system.

*DHIN* employs smart contracts (SC) to manage crucial processes, including for coordinating and rewarding AI training, and for enabling Evaluators to assess patient contributions in a secure, novel proof-of-stake ecosystem. By leveraging these decentralized mechanisms, the framework ensures scalable, sovereign data solutions that respect individual rights and drive technological advancements.

Patients receive cryptographic micropayments into their digital wallets as an incentive to opt into the system. This not only provides immediate value to individuals for their data contributions but also aligns with a long-term roadmap aimed at funding decentralized insurance solutions. This innovative approach could potentially transform the healthcare financing landscape.

By leveraging these decentralized mechanisms, the *DHIN* framework not only addresses the current challenges in health data management but also catalyzes the adoption of scalable, sovereign data solutions. These solutions uphold individual rights, foster technological advancements, and pave the way for new use cases in healthcare. Ultimately, *DHIN* supports the ongoing transition towards decentralized data ownership, promising a more equitable, efficient, and innovative healthcare ecosystem for the future.

# 5 Methodology

## 5.1 Root layer

### 5.1.1 Self-Sovereign Server Technology

In the root layer of the *DHIN* framework, we encounter the foundational technology that enables true data sovereignty: Self-Sovereign Server Technology (SSST). This innovative approach re-defines how individuals, particularly patients, manage their digital identities and health data in a decentralized ecosystem.

At its core, SSST functions as an identity container, combining a user-friendly mobile interface with a continuously connected server. This setup allows patients to securely store and control their health attributes (such as prescription details or lab results), access policies, and transaction receipts [4]. Imagine having a digital vault on your smartphone, where you not only keep your most sensitive health information but also dictate who can access it and when.

Blockchain technology is well-suited for decentralized identity (DID) that does not depend on a centralized root of trust [4]. DID extends blockchain methods to enable a lifelong practical and reliable identifier (e.g., a public/private cryptographic key pair) and attributes linked to that identifier under the self-sovereign control of the individual person [4]. A number of DID systems are emerging, and this technology under the total control of the physician (MD) and the patient can leverage DID to allow for a prescription or equivalent regulated transaction [10].

**A key component of SSST is the use of decentralized identifiers (DIDs) [10]. With DID-based self-sovereign identity:**

- Patients can selectively disclose only necessary health information to providers.
- Providers can authenticate themselves and access patient records with patient permission.
- Health data can be linked to patient DIDs while remaining under patient control.
- Interactions between patients and providers can be securely logged using DIDs.

The white paper "Powering the Physician-Patient Relationship with HIE of One Blockchain Health IT," presented by Adrian Gropper, MD, in 2016, provides an excellent overview of the real-world benefits SSST can offer to both individuals seeking healthcare and the professionals providing it [10]. The major benefit of self-sovereign support technology in this context is the re-decentralization of the trusted relationship between the physician and the patient. The full value of the medical consultation is now available to the two principal parties, with each managing their own policies to provide access to shared resources such as a physician directory or a pharmacy. Additionally, this approach enhances security through diversity, as patients and physician can adopt different self-sovereign technologies to support their respective security, privacy, and economic interests [10].

The beauty of SSST lies in its versatility and security. Each participant in the healthcare ecosystem – be it a patient or a physician – interacts through their own identity container. These containers can be linked to a public blockchain ledger via unique public/private key pairs, ensuring both accessibility and security [4], [10].

For physicians, SSST offers a secure method of authentication. Picture a physician using their smartphone to log into the system, their professional credentials verified through a trusted reputation mechanism – perhaps a digital version of their medical license [10]. This process employs various security measures, including digital signatures, encryption, and even biometric data like fingerprints, all managed through an intuitive user interface [10].

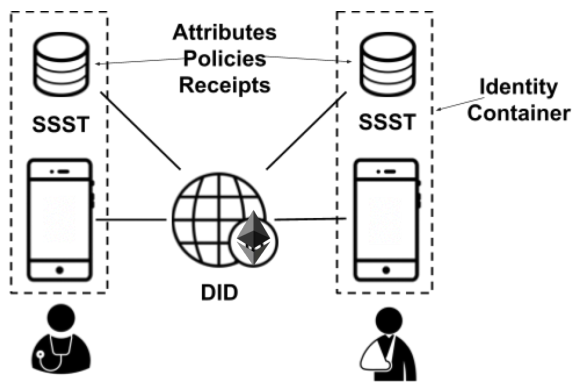

*Figure 1. A transaction between a physician and a patient. Adapted from HIE of One* [10].

Patients, on the other hand, gain unprecedented control over their health records. They can choose which open-source servers to use for storing their personal health information and manage access permissions with ease. It's akin to having a personal health safe, where the patient holds the only key and can grant or revoke access as they see fit.

This system facilitates seamless, secure interactions between patients and physicians. When a physician needs to access a patient's records, they use their sovereign identity to establish a verified relationship with the patient, request access to relevant health data (e.g., via API), and provide care based on comprehensive, up-to-date information [10]. These interactions can be securely logged using DIDs, creating an auditable trail of data access and sharing [4].

The root layer, powered by SSST, lays the groundwork for a healthcare ecosystem where data privacy and individual autonomy are paramount. It transforms the traditional model of centralized health records into a patient-centric system, where individuals are truly in control of their health data. This aligns with the goals of patient-centric, interoperable health IT systems, fostering trust, efficiency, and improved healthcare outcomes [2], [5], [11].

**5.1.2 Personal Health Record**

In the context of the *DHIN* framework, the Personal Health Record (PHR) functions as a Personal Data Store (PDS)—a concept that, while central to this discussion, will not be elaborated on here for brevity. See references [12], [13]. Building upon the foundation of Self-Sovereign Server Technology (SSST), the PHR is a critical component within the *DHIN* ecosystem, enabling patients to manage and share their health information securely and efficiently. Unlike traditional health records, the PHR is designed to be a dynamic, patient-controlled gateway to a lifetime of health information, protected by multiple layers of cryptographic security [10], [14].

At the core of this system is SSST, which acts as a secure bridge connecting healthcare professionals to the patient's health records. With the patient's consent, healthcare professionals can access and update these records through a secure, user-friendly interface [10], [14]. This process is similar to using a specialized, highly secure app, ensuring both ease of use and strong protection of sensitive health data.

A distinguishing feature of the *DHIN* framework is its ability to incorporate and expand upon established principles, such as those discussed in pre-existing works. The framework offers flexibility in managing access permissions by allowing patients to store these permissions either off-chain (on a private, secure server) or on-chain (directly on the blockchain). This approach ensures precise control over who can access their information and under what conditions [15], [16]. This system functions like a digital lock on a medical file cabinet, enabling patients to grant or revoke access to specific records for different healthcare providers as needed.

This level of control raises important considerations regarding the balance between patient autonomy and medical expertise. For example, while patients are generally motivated to maintain accurate records, certain scenarios—such as psychiatric care or pediatrics—may require professional oversight [17], [18], [19]. The *DHIN* framework is designed to accommodate these nuances, enabling customized access policies that respect both patient rights and the need for expert medical guidance.

The technical implementation is streamlined and secure. Healthcare providers access patient records through an API portal, governed by the permissions specified by the patient [20]. This process is akin to having a personal health data concierge, ensuring that only those with the correct credentials and permissions can access sensitive information.

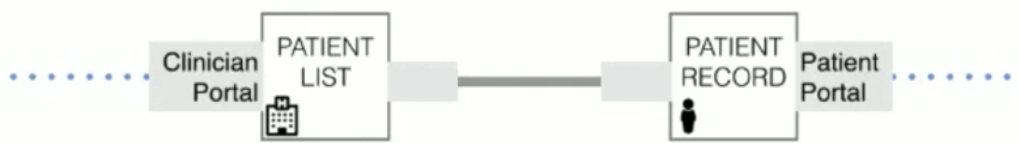

*Figure 2.* API access to personal health record. Adapted from NOSH & HIE of One [20].

This patient-centric approach to health records not only enhances privacy and control but also has the potential to standardize health information formats across different providers. For instance, when moving to a new city, a patient's entire medical history could be instantly accessible to their new healthcare provider without the need for complicated record transfers or redundant testing [20].

Preliminary testing of these concepts has been conducted using open-source PHR systems like NOSH in family medicine settings [19]. Additionally, these ideas have been demonstrated through an integration with Gropper's SSST, providing a demo simulation of a real-life patient encounter between patients and physicians [20]. The principles and applications demonstrated in these settings can be extracted and applied agnostically to any Personal Health Record (PHR) system. This highlights the versatility and adaptability of these concepts within the context of the *DHIN* framework.

In summary, the PHR within the *DHIN* framework empowers patients to move from passive recipients of healthcare to active managers of their health information. This approach fosters greater engagement, accuracy, and continuity in medical care, placing patients at the center of their healthcare journey.

## 5.2. Real-World Layer

**5.2.1 Interaction Layer: Redefining Patient-Doctor Relationships**

The interaction between patients and physicians in the *Decentralized Healthcare Intelligence Network* (*DHIN*) framework represents a significant shift in healthcare provision, leveraging blockchain technology and self-sovereign identities to create a secure, patient-centric ecosystem. The objective is not merely to add a layer of technological sophistication but to fundamentally reshape how healthcare interactions are conducted, ensuring greater transparency, privacy, and patient control.

In the *DHIN* system, physicians establish their professional identity using a cryptographic key pair, functioning as a highly sophisticated, unforgeable digital signature [20]. Unlike traditional systems where credentialing might involve multiple intermediaries, in this framework, the physicians identity is verified by trusted entities such as medical societies or reputable third-party services [20]. This verification process ensures that only qualified professionals can access the system, aligning with the need for trust and integrity in healthcare interactions [20].

**Key Benefits:**

- **Credentialing by Trusted Entities:** Medical societies or established healthcare providers credential medical professionals, offering a trusted setup that patients are familiar with. This integration retains the confidence of patients while empowering doctors to operate within a decentralized framework, potentially recommending cost-effective treatments or referring patients to services outside traditional provider networks [10].
- **Enhanced Professional Autonomy:** Doctors can maintain full control over their practice, free from provider constraints, which can stimulate better market conditions and more personalized patient care [10].

When a doctor needs to access a patient's records, the process is designed to be seamless and secure. Rather than using a conventional username and password, physicians utilize their unique digital identity, which is verified against official medical directories [10]. This method ensures that only authorized professionals can access patient information, preserving both security and privacy.

**Real-World Example:** Imagine a scenario where a doctor prescribes medication or orders tests. Instead of a paper-based or email system prone to forgery and errors, the *DHIN* framework employs a digital identity system where each action by the physician – whether writing a prescription or ordering a test – is digitally signed [10]. This is akin to a digital prescription pad where each prescription is verifiably authentic and tamper-proof.

**Consider the process of filling a prescription [10]:**

- **Digital Prescription Transmission:** The doctor sends a digitally signed prescription directly to the patient's health record.
- **Patient Access and Verification:** At the pharmacy, the patient uses their own digital identity to access the prescription.
- **Pharmacy Verification:** The pharmacy's system verifies both the patient's identity and the doctor's digital signature, ensuring that the prescription is authentic and intended for the patient in question.

This streamlined process eliminates the need for physical prescriptions, reducing the potential for errors and fraud while maintaining a high level of security.

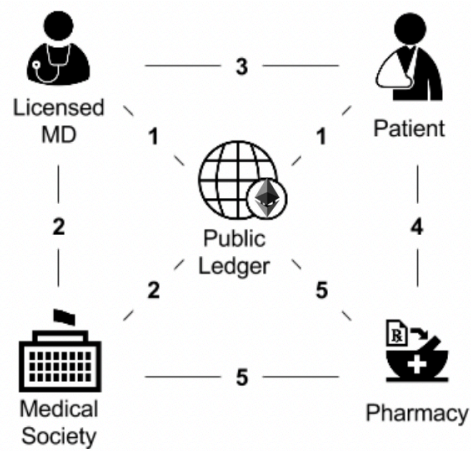

*Figure 3. An example of patient use of a pharmacy service. Adapted from HIE of One* [10].

**Insights from Gropper (2020) [10]:**

- **Credentialing Services:** The Medical Society provides valuable credentialing to its members without risking patient data breaches.
- **Physician Autonomy:** Licensed MDs have total control over their relationships with patients, maximizing the value of their professional licenses.
- **Patient Empowerment:** Patients can choose their pharmacies, preserving privacy and potentially saving money by working directly with their physician.
- **Pharmacy Innovation:** Pharmacies and other healthcare providers can innovate and add value independently of intermediary hospitals and EHR vendors.

- **Support Services:** Suppliers of SSST can offer value-added services to both physicians and patients without patient lock-in or disrupting the institutional trust chain.

The *DHIN* architecture extends beyond prescriptions to all aspects of healthcare, including referrals, imaging services, immunization records, and consent forms. Each interaction is secured, verified, and recorded, contributing to a comprehensive, patient-controlled health record.

**Decentralization and Patient Empowerment:** The decentralization inherent in the *DHIN* framework means that neither patients nor doctors need to rely on large, centralized electronic health record providers. Instead, each participant in the healthcare ecosystem – whether doctor, patient, or healthcare service provider – can interact directly and securely, with the patient maintaining central control.

In essence, the *DHIN* framework draws upon and extends the principles laid out by Gropper [10], particularly in establishing sovereign data stores that facilitate secure and efficient healthcare exchanges. While not limited to Gropper's work, these technologies are integral to the *DHIN* framework, embedding trust in every interaction, prioritizing privacy, and empowering patients to maintain true ownership of their healthcare journey.

### 5.2.2 Medical AI Devices

The *DHIN* framework introduces a theoretical model for integrating Medical AI Devices into a decentralized healthcare system. These systems are defined in this paper as entities that handle health data—whether by storing, retrieving, generating, sending, or processing it, including tools such as triage systems, diagnostic instruments, and AI decision support systems.

Currently, implementing AI systems in clinical environments faces major challenges [21]. A significant issue is the fragmentation of health data across various provider systems, which complicates the development and deployment of effective AI tools [4], [15], [21], [22], [23], [24]. These tools typically need accurate and up-to-date data from a single, unified source.

The proposed framework addresses this challenge with an approach that might seem ironic at first: it suggests centralizing all patient health data in one location. However, this centralization is not in the traditional sense of a single, centralized database. Instead, it centralizes data at the point where it is most relevant—the patient's own Personal Health Record (PHR). This means that while the data is distributed across various sources, it is centralized in the hands of the individual it concerns across their lifetime. This method resolves disputes over data access between healthcare providers and ensures that AI-generated outputs are recorded in the patient-owned PHR. A more complete PHR with information held over an individual's lifetime will enable more comprehensive information, thus enhancing the precision and accuracy of AI tools (e.g., decision support) as they work with more complete data about the patient.

In this model, Medical AI Devices would each be assigned their own Decentralized Identifiers (DIDs). These DIDs function similarly to unique digital IDs for people or organizations, allowing devices to interact directly with the patient's PHR [4]. This setup enables devices to read from and write to the PHR as needed.

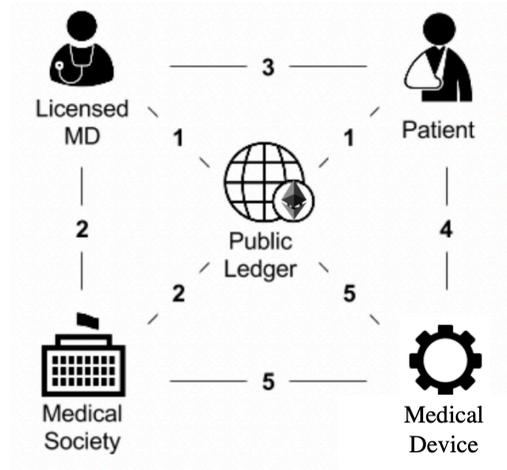

**Figure 4.** *An example of patient use of a Medical AI Device* [4]. *Adapted from HIE of One* [10].

Each system makes use of the same authentication architectures as physicians to attribute licensing/credentials of their system (e.g., FDA/CE, trusted third-party), as shown in **Figure 4**. Regardless of authentication, the incorporation of DIDs can reduce the manual effort of device labeling and provide a safer use of medical devices using their blockchain-based identities [25], although many open-source systems do not require licensing and authentication in practice. An open-source Medical AI Device is ideal for agile prototyping among physicians, patients, developers, and other key stakeholders, facilitating the adaptation, training, and deployment of AI tools.

Medical devices today are capable of storing control software and data, making them suitable for using DIDs [4]. By consolidating patient health data into a PHR, these devices can more effectively coordinate and manage health information. Medical AI Devices would analyze data from the PHR to enhance healthcare coordination.

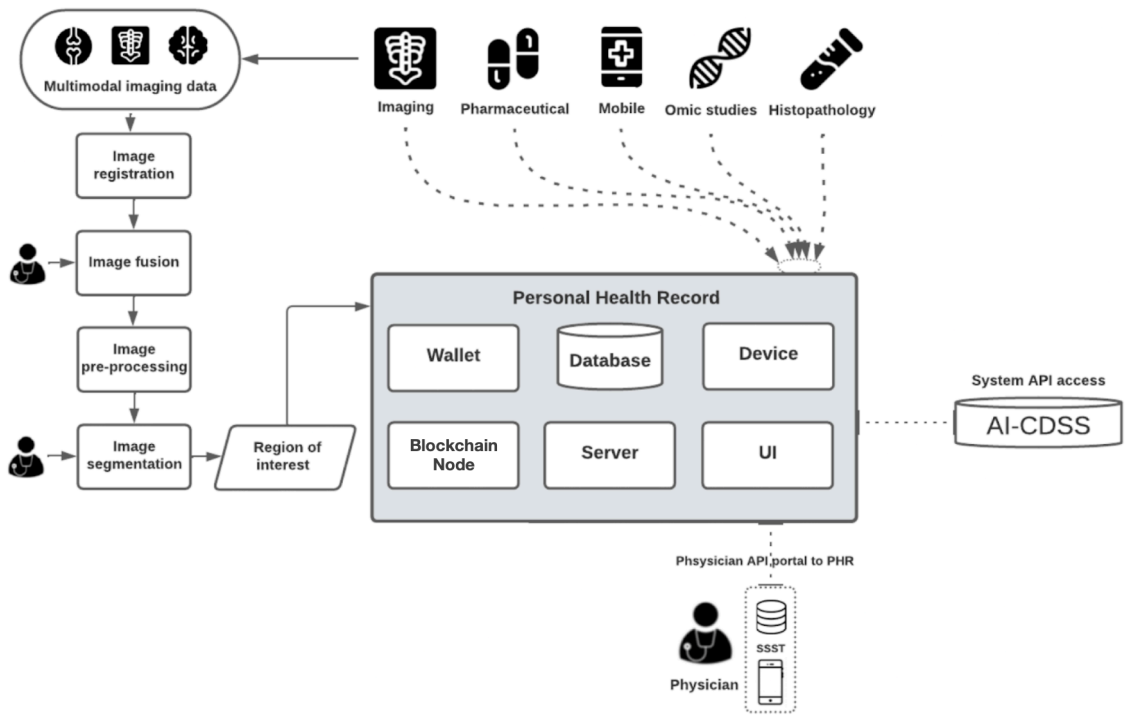

*Figure 5. Interactions amongst physicians and supporting systems that read and write directly into the personal health record are mediated by the blockchain.*

Medical AI Devices working with physicians would use the PHR to record their outputs, promoting better collaboration between physicians and AI tools. This approach aids in making informed healthcare decisions and enhances the accuracy of specific tasks in the healthcare process, such as diagnostic tools or mobile health devices.

The *DHIN* framework envisions Medical AI Devices running algorithms on the data stored in the PHR. This data could include details entered by physicians (e.g., symptoms, medical history, physical examination results) and additional information from sources like imaging or mobile health devices. The paper acknowledges that this is a relatively new area of research and calls for further study into integrating Medical AI Devices into clinical workflows using patient-owned PHR systems. This includes ensuring that these devices access accurate health data, assigning DIDs to Medical AI Devices, and effectively integrating AI with medical practice.

To implement this framework, a decentralized federated learning (FL) environment is essential. This environment would allow AI tools to develop using patient-owned health data while maintaining patient control over their information.

In summary, this theoretical framework sets the stage for the next layer of the system—the intelligence layer—focused on overcoming the challenges of deploying AI in a decentralized healthcare ecosystem.

### 5.3 Intelligence Layer

The *Decentralized Health Intelligence Network (DHIN)* directly implements the *Decentralized Intelligence Network (DIN)* protocol [3] as its core intelligence layer. This section briefly outlines the key components of the *DIN* protocol and discusses their specific implications and applications within the context of *DHIN's* vision for sovereign healthcare. For a comprehensive explanation of the protocol's technical details, readers are strongly encouraged to refer to the original *DIN* paper [3].

#### 5.3.1 Overview of DIN Protocol in DHIN

The *DIN* protocol forms the foundational architecture for *DHIN's* intelligence layer. *DHIN* adopts the protocol in its entirety, applying its principles and mechanisms to enable sovereign healthcare.

**Key components of the *DIN* protocol implemented in *DHIN* include [3]:**

1. Decentralized public blockchain infrastructure
2. Federated Learning (FL) architecture
3. Personal Data Stores (PDS), utilized as Personal Health Records (PHRs) in *DHIN*
4. Smart contracts for coordination and rewards FL processes
5. Off-chain decentralized file storage (e.g., IPFS)
6. Secure aggregation and evaluation processes

At its core, this system uses cryptographic micropayments as a reward mechanism for patients who contribute their health data to AI development. Specifically, micropayments are used by Model Owners to access on-chain smart contracts that coordinate the federated learning process with patients' personal health records (PHRs), which serve as personal data stores (PDS).

This innovative use of cryptographic micropayments not only incentivizes patient participation but also ensures that data contributors are fairly compensated for the value their data brings to AI development. By integrating smart contracts, the system automates the coordination of federated learning, enabling secure, privacy-preserving updates to AI models without exposing sensitive personal health records. This approach transforms healthcare data management, turning it into a decentralized, patient-centric process that empowers individuals to maintain sovereignty over their data while actively participating in the advancement of medical research and personalized care.

Furthermore, the protocol implements a novel NFT Proof of Stake (PoS) mechanism for Evaluators, combining innovative evaluation techniques with Harberger taxation and Partial Common Ownership (PCO) principles. This approach not only secures the network by requiring Evaluators to stake NFTs but also contributes to public goods funding through a taxation system. The mechanism incentivizes high-quality evaluations and active participation while preventing monopolistic behavior [3]. By linking Evaluator rewards to their stake and performance, the protocol ensures a fair, efficient, and secure system for decentralized machine learning model evaluation and improvement.

### 5.3.2 Decentralized Insurance Solution

The embedded *DIN* protocol enables the *DHIN* framework to introduce an innovative approach to healthcare financing by utilizing rewards received from opting into the FL decentralized insurance solution, creating a self-sustaining cycle of value in healthcare.

Imagine receiving a small digital payment each time your health information helps train an AI model to improve healthcare outcomes. These micropayments are deposited directly into patients' digital wallets (e.g., Taho [26]), seamlessly and securely.

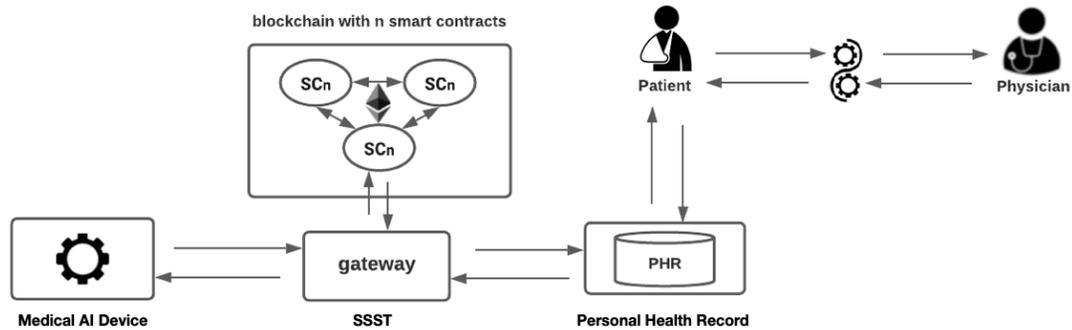

*Figure 6.* The rewards mechanism of cryptographic insurance in a learning health system.

**Here's how the cycle works:**

1. Patients and physician interact as usual, with health data recorded in the patient's Personal Health Record (PHR).
2. Model Owners (i.e., Medical AI Devic Developer) use cryptographic payments to access on-chain smart contracts, allowing them to train their models on patients' health data through federated learning protocols.
3. Patients receive these micropayments in their digital wallets when they successfully contribute to an AI training round.
4. Patients can then use these rewards to purchase decentralized insurance premiums, which in turn fund or subsidize their healthcare provision.

This cyclical mechanism creates a learning health system where value continuously circulates. New health technologies, embedded in AI models, inform and improve healthcare delivery. As patients use their rewards to fund their care, they generate more valuable health data, which then feeds back into AI development.

Importantly, this system preserves patient sovereignty over their data while providing AI tools permissionless access to up-to-date health information. It's a win-win scenario: patients benefit financially from their data, and AI systems get access to high-quality, diverse datasets.

The system also addresses a key challenge in healthcare: funding for services and treatments. By creating a fungible asset (a cryptographic micropayment i.e., a stablecoin) that can be used for insurance premiums, the framework establishes a sustainable funding model for decentralized healthcare.

In federated learning (FL) with non-homogenous data sources such as health data, data quantity and quality are the most valuable contribution to enhancing the accuracy of training machine learning algorithms [27]. Interestingly, this system may naturally benefit those with chronic or ongoing health conditions. These patients, who typically have more extensive health records due to frequent care, are likely to receive greater rewards in the learning process [27]. This aligns well with their potentially higher insurance needs. Furthermore, learning on health data is not limited to one occasion or use-case, but multiple occasions and use-cases.

Looking to the future, this framework lays the groundwork for more efficient insurance claim verification. By leveraging blockchain technology, the system could provide a reliable, immutable source of information for verifying insurance credentials and claims [4]. Decentralized insurance protocols are well suited to enhance the functions of more cost-effective and reliable coverage schemes [4].

The long-term vision is ambitious yet promising: a scalable system that could potentially reduce healthcare insurance costs, lower barriers to healthcare provision, and ultimately increase access to quality healthcare for all.

In essence, the *DHIN's* decentralized insurance solution represents a paradigm shift in healthcare financing, aligning patient interests, technological advancement, and healthcare provision in a novel, self-sustaining ecosystem.

## 6. DHIN-Specific Applications

The integration of the *DIN* protocol within the *Decentralized Health Intelligence Network (DHIN)* introduces transformative implications for sovereign healthcare systems:

1. **Sovereign Health Data Management**:
    - Personal Health Records (PHRs) in *DHIN* act as patient-controlled, sovereign health data repositories.
    - Patients maintain granular control over their health data, deciding which studies, applications, or healthcare providers can access specific parts of their records.
    - Importantly, no data exchanges are necessary; all learning and data processing occur within the patient's personal device or secure enclave.
2. **Federated Learning for Personalized Medicine**:
    - Facilitates the development of AI models that learn from diverse datasets that are geographically unbounded, while preserving individual privacy.
    - Supports the creation of personalized treatment plans based on comprehensive, real-world data, unbounded by institutional limitations.
    - Enhances rare disease research by leveraging geographically unbounded and distributed patient data while upholding strict data sovereignty.
3. **Smart Contracts for Decentralized Clinical Trials**:
    - Manage patient consent and participation in medical studies through opt-in smart contracts, providing patients with fine-grained control over the specific data being used for learning, all at their discretion.
    - Automate compliance with healthcare regulations (e.g., HIPAA, GDPR) at the individual participant level.
    - Enable truly decentralized clinical trials, transcending geographical and institutional barriers, and expanding patient participation.
4. **Secure Aggregation for Population Health Insights**:
    - Enables the development of robust population health models without centralizing or exposing individual patient data.
    - Facilitates early detection of health trends and potential outbreaks on a global scale through secure data aggregation.
5. **Decentralized Auditing and Reward Distribution**:
    - Incentivizes patient participation through transparent and fair reward mechanisms without introducing a third party broker (i.e., all rewards pass to patients).
    - Distributes rewards directly to patients' digital wallets, potentially in the form of stablecoin cryptocurrency (e.g., USDC, RAI, etc.).
    - Lays the groundwork for decentralized health insurance solutions, enabling wider access to healthcare services and creating a self-financing healthcare ecosystem. This approach unlocks previously unutilized value that can complement existing public and private healthcare funding options.
6. **Incentive Mechanisms for Global Health Research**:
    - Encourages global patient participation in medical studies by offering direct, secure incentives.
    - Promotes collaboration across healthcare institutions, researchers, and individual patients worldwide, fostering innovation in healthcare delivery.

# 7. Conclusion and Future Directions

The *Decentralized Health Intelligence Network (DHIN)* stands as a pioneering extension of the *Decentralized Intelligence Network* (*DIN*), offering a transformative approach to healthcare data management. By building on *DIN's* foundational principles, *DHIN* specifically addresses the challenges inherent in the healthcare sector through the integration of Personal Health Records (PHRs), decentralized federated learning, and a trustless rewards system. This framework empowers patients with full control over their health data, securely stored within their PHRs, and facilitates seamless interactions with healthcare providers and researchers via straightforward opt-in or opt-out mechanisms.

The use of *DIN's* scalable federated learning protocol on a public blockchain ensures that patient data remains within the PHR, never leaving its secure environment. Instead, only model parameter updates are shared, using advanced privacy-preserving techniques to safeguard data throughout the process. This approach effectively addresses the issue of data silos, enabling large-scale, collaborative healthcare research while upholding strict individual privacy standards. Broader adoption of *DHIN* allows the federated learning protocol to operate across various geographical regions, expanding the data pool beyond isolated, patient-specific sources. This enables AI tools to achieve greater accuracy, especially for rare diseases, by leveraging a more comprehensive dataset that transcends individual patient silos.

**Looking to the future, *DHIN* is poised to significantly enhance its impact through several key directions:**

- **Integration with Personal Health Devices and Wearables:** Expanding the network to include sovereign devices and hardware, where the data collected is owned by the individual rather than a specific company, further strengthens patient control over their health data.
- **Development of Patient-Centric Interfaces:** Creating user-friendly interfaces that allow patients to effortlessly opt into federated learning protocols and manage their participation in studies, ensuring greater patient engagement and accessibility.
- **Global Decentralized Health Data Marketplaces:** Facilitating the creation of marketplaces where individuals can securely and privately monetize their health data, driving global collaboration in medical research.

The *DIN* protocol plays a crucial role in enabling these future directions by facilitating data monetization and supporting learning on sovereign data. **This decentralized approach brings about significant cost reductions for healthcare institutions:**

- **Elimination of Centralized Data Storage:** By allowing data to remain within personal health records, *DHIN* removes the need for centralized data repositories, reducing the costs associated with maintaining and securing large-scale data storage systems.
- **Mitigation of Risks and Costs Associated with Data Breaches:** Decentralizing data storage significantly reduces the risks of large-scale data breaches, which can be financially devastating for institutions. By keeping data decentralized and secure within PHRs, the network minimizes these risks and the associated costs.
- **Reduction of IT Infrastructure Requirements:** Since data does not need to be centrally stored or managed, healthcare providers can reduce their IT infrastructure requirements, leading to lower operational costs and more efficient use of resources.

In summary, *DHIN* provides a robust foundation for a truly decentralized, privacy-preserving, and globally collaborative health intelligence network. This innovative approach has the potential to revolutionize medical research by enabling sovereign healthcare practices and fostering direct patient involvement on a global scale, all while maintaining strict data sovereignty and reducing institutional costs. We invite researchers, practitioners, and stakeholders to engage with the *DHIN* framework. By collaborating, we can advance scalable, sovereign data solutions that enhance healthcare technology, respect individual rights, and build a more equitable, efficient, and innovative healthcare system that benefits all participants.